\documentclass[conference]{IEEEtran}
\IEEEoverridecommandlockouts
\usepackage{cite}
\usepackage{amsmath,amssymb,amsfonts, bbm}
\usepackage{algorithmic}
\usepackage{graphicx}
\usepackage{textcomp}
\usepackage{xcolor}
\def\BibTeX{{\rm B\kern-.05em{\sc i\kern-.025em b}\kern-.08em
    T\kern-.1667em\lower.7ex\hbox{E}\kern-.125emX}}

\usepackage{amsmath,amssymb,amsfonts,bbm,bm,mathrsfs}
\usepackage{color}
\usepackage{algorithmic}
\usepackage{graphicx}
\usepackage{textcomp}
\usepackage{xcolor}
\usepackage{amsthm}

\usepackage[ruled,vlined,linesnumbered]{algorithm2e}

\usepackage[mode=tex]{standalone} % mode=image|tex, mode=build, buildmissing

\usepackage{tikz}

\usepackage{chronosys}
\usepackage{xspace}
\usetikzlibrary{shapes,arrows,calc,positioning,fit,automata}
\usetikzlibrary{decorations.markings}
\usetikzlibrary{overlay-beamer-styles}
\usepackage{fontawesome}

\usepackage{pgfplots}
\usepackage{pgfplotstable}
\usepackage{filecontents}

\usepackage{caption,subcaption}

\newcommand{\vect}[1]{\ensuremath{\mathbf{#1}}}

\newcommand{\myBlue}{blue!80!black}
\newcommand{\myGreen}{green!60!black}
\newcommand{\myRed}{red!80!black}

\newcommand{\Ndata}{N_\text{data}}
\newcommand{\Nsc}{N_\text{sc}}
\newcommand{\Nse}{N_\text{se}}
\newcommand{\Nfft}{N_\text{FFT}}

\newcommand{\EBW}{\text{EBW}}
\newcommand{\ER}{\text{ER}}

\newcommand{\SER}{\text{SER}}
\newcommand{\PAPR}{\text{PAPR}}

\definecolor{NYUviolet}{HTML}{57068c} 	% official NYU violet
\definecolor{NYUlight}{HTML}{8900e1} 	% NYU light violet
\definecolor{NYUdark}{HTML}{330662} 	% NYU dark violet
\definecolor{NYUnight}{HTML}{220337} 	% NYU darker violet

\tikzstyle{block}=[rectangle,draw,very thick,fill=white,align=center,font=\Large]
\tikzstyle{edge} = [draw,very thick,->,-triangle 45]
\tikzstyle{plateBlue} = [draw=\myBlue, shape=rectangle, rounded corners=0.5ex, ultra thick,
minimum width=3.1cm, text=\myBlue, text width=3.1cm, align=right, inner sep=10pt, inner ysep=10pt,append after command={node[font=\bf,text=\myBlue,above left= 3pt of \tikzlastnode.south east] {#1}}]

\tikzstyle{plateRed} = [draw=\myRed, shape=rectangle, rounded corners=0.5ex, ultra thick,
minimum width=3.1cm, text=\myRed, text width=3.1cm, align=right, inner sep=10pt, inner ysep=10pt,append after command={node[font=\bf\LARGE,text=\myRed,above= 3pt of \tikzlastnode.south] {#1}}]

\tikzstyle{plateGreen} = [draw=\myGreen, shape=rectangle, rounded corners=0.5ex, ultra thick,
minimum width=3.1cm, text=green!80!black, text width=3.1cm, align=right, inner sep=10pt, inner ysep=10pt,append after command={node[font=\bf\LARGE,text=\myGreen,above= 3pt of \tikzlastnode.south] {#1}}]

\tikzstyle{plate} = [draw, shape=rectangle, rounded corners=0.5ex, ultra thick,
minimum width=3.1cm,  text width=3.1cm, align=right, inner sep=10pt, inner ysep=10pt,append after command={node[font=\bf\Large,above left= 3pt of \tikzlastnode.south east] {#1}}]

\tikzstyle{plateSmall} = [draw, shape=rectangle, rounded corners=0.5ex, ultra thick,
minimum width=3.1cm,  text width=3.1cm, align=right, inner sep=10pt, inner ysep=10pt,append after command={node[font=\bf,above= 3pt of \tikzlastnode.south] {#1}}]

\usepackage[prefix=]{xcolor-material}
\pgfmathdeclarerandomlist{material}{%
	{Red}{Blue}{Green}}
\tikzset{%
	half clip/.code={
		\clip (0, -256) rectangle (256, 256);
	},
	color/.code=\colorlet{fill color}{#1},
	color alias/.code args={#1 as #2}{\colorlet{#1}{#2}},
	colors alias/.style={color alias/.list/.expanded={#1}},
	execute/.code={#1},
	on left/.style={.. on left/.style={#1}},
	on right/.style={.. on right/.style={#1}},
	split/.style args={#1 and #2}{
		on left ={color alias=fill color as #1},
		on right={color alias=fill color as #2, half clip}
	}
}
\newcommand\reflect[2][]{%
	\begin{scope}[#1]\foreach \side in {-1, 1}{\begin{scope}
				\ifnum\side=-1 \tikzset{.. on left/.try}\else\tikzset{.. on right/.try}\fi
				\begin{scope}[xscale=\side]#2\end{scope}
\end{scope}}\end{scope}}

\tikzset{%
	cat/.pic={
		\tikzset{x=1.5cm/5,y=1.5cm/5,shift={(0,-1/3)}}
		\useasboundingbox (-1,-1) (1,2);
		\fill [BlueGrey900] (0,-2)
		.. controls ++(180:3) and ++(0:5/4) .. (-2,0)
		arc (270:90:1/5)
		.. controls ++(0:2) and ++(180:11/4) .. (0,-2+2/5);
		\foreach \i in {-1,1}
		\scoped[shift={(1/2*\i,9/4)}, rotate=45*\i]{
			\clip [overlay] (0, 5/9) ellipse [radius=8/9];
			\clip [overlay] (0,-5/9) ellipse [radius=8/9];
			\fill [BlueGrey900] ellipse [radius=1];
			\clip [overlay] (0, 7/9) ellipse [radius=10/11];
			\clip [overlay] (0,-7/9) ellipse [radius=10/11];
			\fill [Purple100] ellipse [radius=1];
		};
		\fill [BlueGrey900] ellipse [x radius=3/4, y radius=2];
		\fill [BlueGrey100] ellipse [x radius=1/3, y radius=1];
		\fill [BlueGrey900]
		(0,15/8) ellipse [x radius=1, y radius=5/6]
		(0, 8/6) ellipse [x radius=1/2, y radius=1/2]
		{[shift={(-1/2,-2)}, rotate= 10]  ellipse [x radius=1/3, y radius=5/4]}
		{[shift={( 1/2,-2)}, rotate=-10] ellipse [x radius=1/3, y radius=5/4]};
		\fill [BlueGrey500]
		(-1/9,11/8) ellipse [x radius=1/5, y radius=1/5]
		( 1/9,11/8) ellipse [x radius=1/5, y radius=1/5];
		\fill [Purple100]
		(0,12/8)     ellipse [x radius=1/10, y radius=1/5]
		(0,12/8+1/9) ellipse [x radius=1/5 , y radius=1/10];
		\foreach \i in {-1,1}
		\scoped[shift={(1/2*\i,2)}, rotate=35*\i]{
			\clip [overlay] (0, 1/7) ellipse [radius=2/7];
			\clip [overlay] (0,-1/7) ellipse [radius=2/7];
			\fill [Yellow50] ellipse [radius=1];
		};
		\scoped{
			\clip (-1,-2) rectangle ++(2,1);
			\fill [BlueGrey900] (0,-2) ellipse [radius=1/2];
			\fill [Grey100]
			(-1/2,-2) ellipse [x radius=1/3, y radius=1/4]
			( 1/2,-2) ellipse [x radius=1/3, y radius=1/4];
		};
		\foreach \i in {-1,1}
		\foreach \j in {-1,0,1}
		\fill [Grey100, shift={(0,11/8)}, xscale=\i, rotate=\j*15,
		shift=(0:1/2)]
		ellipse [x radius=1/3, y radius=1/64];
	},
dog/.pic={
	\begin{scope}[x=1.5cm/480,y=1.5cm/480]
		\useasboundingbox (-256, -256) (256, 256);
		\reflect[split=Brown400 and Brown500]{
			\fill [fill color] (0,-64) ellipse [x radius=160, y radius=144];
			\fill [fill color] (0, 32) ellipse [x radius=128, y radius=112];
			\fill [fill color] (32,96)
			.. controls ++( 75:128) and ++(105:128) .. ++(192,  0)
			.. controls ++(285: 96) and ++(285: 96) .. ++(-80,-32)
			.. controls ++(105: 64) and ++( 75: 32) .. cycle;
		}
		\reflect[split={Grey100 and Grey200}]{
			\clip (0,-64) ellipse [x radius=160, y radius=144];
			\fill [fill color](0,-224) 
			.. controls ++(  0:64) and ++(270:64) .. ++(96,64)
			.. controls ++( 90:64) and ++(270:64) .. ++(-96,192)
			.. controls ++(270:64) and ++( 90:64) .. ++(-96,-192)
			.. controls ++(270:64) and ++(180:64) .. cycle;
		}
		\reflect[split={Pink100 and Pink200}]{
			\fill [fill color](0,-192) ellipse [x radius=28, y radius=32];
		}
		\reflect[split={BlueGrey800 and BlueGrey900}]{
			\fill [fill color](0,-144) 
			.. controls ++(  0:20) and ++(315:20) .. ++( 40,64)
			.. controls ++(135:20) and ++( 45:20) .. ++(-80, 0)
			.. controls ++(225:20) and ++(180:20) .. cycle;
			\fill [BlueGrey900] (56, 0) circle [radius=20];
			\fill [fill color] (-8,-112)
			-- ++(16,0) -- ++(0,-32) arc (180:360:24)
			arc (180:0:8) arc (360:180:40);
		}
\end{scope}}
}

\tikzset{
	o/.style={
		shorten >=#1,
		decoration={
			markings,
			mark={
				at position 1
				with {
					\draw circle [radius=#1];
				}
			}
		},
		postaction=decorate
	},
	o/.default=2pt
}

\tikzset{naming/.style={align=center,font=\large}}
\tikzset{antenna/.style={insert path={-- coordinate (ant#1) ++(0,0.25) -- +(135:0.25) + (0,0) -- +(45:0.25)}}}
\tikzset{station/.style={naming,draw,shape=dart,shape border rotate=90, minimum width=15mm, minimum height=30mm,outer sep=0pt,inner sep=3pt}}
\tikzset{stationPoster/.style={naming,draw,shape=dart,shape border rotate=90, minimum width=20mm, minimum height=40mm,outer sep=0pt,inner sep=3pt}}
\tikzset{mobile/.style={naming,draw,shape=rectangle,minimum width=12mm,minimum height=6mm, outer sep=0pt,inner sep=3pt}}
\tikzset{radiation/.style={{decorate,decoration={expanding waves,angle=90,segment length=4pt}}}}

\begin{document}

\title{
    Learned Pulse Shaping Design for\\ PAPR Reduction in DFT-s-OFDM
    \thanks{
        This work was done in part while F. Carpi was an intern at Samsung Research America. 
        The work of F.~Carpi, S.~Garg, and E.~Erkip was supported in part by the NYU WIRELESS Industrial Affiliates Program, and by the NSF grants \#1925079 and \#2003182.
        Please send correspondence to \texttt{fabrizio.carpi@nyu.edu}.
    }
}

\author{
    \IEEEauthorblockN{
        Fabrizio Carpi\IEEEauthorrefmark{1},  
        Soheil Rostami\IEEEauthorrefmark{2},
        Joonyoung Cho\IEEEauthorrefmark{2},
        Siddharth Garg\IEEEauthorrefmark{1},
        Elza Erkip\IEEEauthorrefmark{1},
        Charlie Jianzhong Zhang\IEEEauthorrefmark{2}
    }
    \IEEEauthorblockA{
        \IEEEauthorrefmark{1}Department of Electrical and Computer Engineering, New York University, Brooklyn, NY\\
        \IEEEauthorrefmark{2}Standards and Mobility Innovation, Samsung Research America, USA
    }
}

\maketitle

\begin{abstract}

High peak-to-average power ratio (PAPR) is one of the main factors limiting cell coverage for cellular systems, especially in the uplink direction.
Discrete Fourier transform spread orthogonal frequency-domain multiplexing (DFT-s-OFDM) with spectrally-extended frequency-domain spectrum shaping (FDSS) is one of the efficient techniques deployed to lower the PAPR of the uplink waveforms.
In this work, we propose a machine learning-based framework to determine the FDSS filter, optimizing a tradeoff between the symbol error rate (SER), the PAPR, and the spectral flatness requirements. 
Our end-to-end optimization framework considers multiple important design constraints, including the Nyquist zero-ISI (inter-symbol interference) condition.
The numerical results show that learned FDSS filters lower the PAPR compared to conventional baselines, with minimal SER degradation.
Tuning the parameters of the optimization also helps us understand the fundamental limitations and characteristics of the FDSS filters for PAPR reduction.

\end{abstract}


%%%%%%%%%%%%%%%%%%%%%
%%%%%%%%%%%%%%%%%%%%%
\section{Introduction}
\label{sec:introduction}

    Orthogonal Frequency Division Multiplexing (OFDM) is a cornerstone technology in modern wireless communication systems, renowned for its ability to efficiently transmit data in high-rate applications and its robustness against multipath propagation. 
    One of the inherent challenges in OFDM systems is the high peak-to-average power ratio (PAPR) of the transmitted signals. 
    This high PAPR demands linear amplification with considerable power back-off (to avoid saturation in power amplifiers), reducing the overall power efficiency and increasing the risk of nonlinear distortion in the transmitters.

    The 3rd Generation Partnership Project (3GPP) has been studying power-domain techniques targeting PAPR reduction for coverage enhacement~\cite{3GPP}.
    Discrete Fourier transform spread OFDM (DFT-s-OFDM) has been employed for 4G/5G cellular uplink to address the PAPR issues of OFDM and to improve coverage~\cite{Book-LTE}.
    In DFT-s-OFDM, the application of a DFT spreading operation prior to subcarrier mapping to IDFT (inverse DFT) results in a single carrier waveform, effectively reducing the PAPR of the transmitted signal. 
    This technique significantly improves power amplifier efficiency and reduces the risk of distortion.

    Several methods for PAPR reduction have been proposed in the past, acting on different blocks of the transmission chain~\cite{survey-PAPR}.
    Different pulse shaping techniques have been investigated in order to obtain low-PAPR waveforms for the vanilla DFT-s-OFDM architecture~\cite{PAPR-SCFDMA, PAPR-combinePulses, PAPR-probabilisticPulseShaping}.  
    Previous works~\cite{Kim-2018-DL, Wang-2021-DL} also used deep learning methods to mitigate PAPR effects, by including additional neural network processing blocks within the traditional OFDM end-to-end architecture. 
    In this work, we focus on the DFT-s-OFDM with spectral extension (SE) followed by the frequency-domain spectrum shaping (FDSS) filter. 
    The goal is to design an FDSS filter that reduces PAPR, at the expense of the excess bandwidth introduced with the spectral extension operation.
    This scheme provides effective pulse shaping in frequency domain, and it is one of the outstanding approaches that 3GPP has investigated for coverage enhancement~\cite{3GPP}.
    While DFT-s-OFDM offers notable advantages, optimizing the FDSS design of the pulse shaping filter presents a unique challenge~\cite{EnhancedUplinkCoverage} and it is often performed in a heuristic fashion. 
    Traditionally, pulse shaping filters have been defined using well-established mathematical functions, such as the raised cosine~\cite{Nyquist-1928} or the exponential~\cite{BetterThanNyquist}, offering limited flexibility in tailoring the filter to specific design requirements.
    Later works have explored parametric construction for the pulse shaping filters~\cite{ParametricConstruction}, analyzing the combination of hyperbolic functions~\cite{ImprovedNyquist, NovelTwoParams}.
    However, none of these filters have been specifically designed for PAPR reduction; the main performance metrics that were considered are sensitivity to timing errors, maximum distortion in eye diagrams, and amplitude decay of the pulse in time domain.

    This paper introduces a novel approach to achieve low PAPR for the DFT-s-OFDM systems by designing FDSS filters with machine learning-based methods. 
    We propose a data-driven end-to-end framework to learn FDSS filters that optimize a tradeoff between three components: (i) the PAPR of the transmitted waveform; (ii) the symbol error rate (SER) degradation due to inter-symbol interference (ISI) possibly introduced by the FDSS filters; (iii) the shape of the resulting FDSS filters in frequency domain.
    The proposed loss function contains the weighted combination of these three abovementioned components; by tuning the importance (weight) of each component, one can obtain filters for different tradeoff requirements. 
    The learned filter taps are modeled with a polynomial approximation which produces \emph{smooth} FDSS filter functions that can be reused for different system configurations.
    Moreover, we also formulate constrained FDSS designs that can capture filter requirements such as flat passband response and Nyquist zero-ISI condition.

%%%%%%%%%%%%%%%%%%%%%
%%%%%%%%%%%%%%%%%%%%%
\section{System Model}
\label{sec:system-model}

    \begin{figure*}[t]
        \centering
        \begin{minipage}[b]{.6\textwidth}
            \begin{tikzpicture}
    \def\dhor{4}
    \def\dvert{5.5}

    % TX Blocks
    \node (s) at (-0.75*\dhor,0) {};
    \node[block, inner sep=5pt, align=center] (mod) at (0,0) {Modulator};
    \node[block, inner sep=5pt, align=center] (DFT) at (\dhor,0) {DFT\\$\Ndata$};
    \node[block, shape=trapezium, rotate=90, inner sep=5pt, align=center] (SE) at (2*\dhor,0) {\rotatebox{-90}{SE}};
    \node[block, draw=\myRed, text=\myRed, inner sep=5pt, align=center] (FDSS) at (3*\dhor,0) {(Learned)\\FDSS\\$\vect{F}$};
    \node[block, inner sep=5pt, align=center] (IFFT) at (4*\dhor,0) {iFFT\\$\Nfft$};

    % Channel
    \node[block, shape=ellipse, inner sep=5pt, align=center] (C-AWGN) at (4.5*\dhor,-0.5*\dvert) {$\mathbb{C}$-AWGN};

    % RX Blocks
    \node[block, inner sep=5pt, align=center] (FFT) at (4*\dhor,-\dvert) {FFT\\$\Nfft$};
    \node[block, draw=\myRed, text=\myRed, inner sep=5pt, align=center] (IFDSS) at (3*\dhor,-\dvert) {Matched\\FDSS\\$\vect{F}^*$};
    \node[block, shape=trapezium, rotate=90, inner sep=5pt, align=center] (ISE) at (2.25*\dhor,-\dvert) {\rotatebox{-90}{iSE}};
    \node[block, inner sep=5pt, align=center, font=\LARGE] (normaliz) at (1.75*\dhor,-\dvert) {$\frac{1}{|\vect{F}|^2}$};
    \node[block, inner sep=5pt, align=center] (IDFT) at (\dhor,-\dvert) {IDFT\\$\Ndata$};
    \node[block, inner sep=5pt, align=center] (demod) at (0,-\dvert) {Demod.};
    \node (s-hat) at (-0.75*\dhor,-\dvert) {};

    % Connections
    \path[edge] (s) --node[above=0.1,font=\Large]{$\vect{s}$} (mod);
    \path[edge] (mod) --node[above=0.1,font=\Large]{$\vect{x}$} (DFT);
    \path[edge] (DFT) --node[above=0.1,font=\Large]{$\vect{X}$} (SE);
    \path[edge] (SE) --node[above=0.1,font=\Large]{$\vect{X}^\text{ext}$} (FDSS);
    \path[edge] (FDSS) --node[above=0.1,font=\Large]{$\vect{\tilde{X}}$} (IFFT);
    \path[edge] (IFFT) -| node[above=0.1,font=\Large]{$\vect{\tilde{x}}$} (C-AWGN);

    \path[edge] (C-AWGN) |-node[below=0.1,font=\Large]{$\vect{\tilde{y}}$} (FFT);
    \path[edge] (FFT) --node[above=0.1,font=\Large]{$\vect{\tilde{Y}}$} (IFDSS);
    \path[edge] (IFDSS) --node[above=0.1,font=\Large]{$\vect{{R}}$} (ISE);
    \path[edge] (ISE) --node[above=0.1,font=\Large]{$\vect{{T}}$} (normaliz);
    \path[edge] (normaliz) --node[above=0.1,font=\Large]{$\vect{\tilde{Y}}$} (IDFT);
    \path[edge] (IDFT) --node[below=0.1,font=\Large]{$\vect{{y}}$} (demod);
    \path[edge] (demod) --node[below=0.1,font=\Large]{$\vect{\hat{s}}$} (s-hat);

    %%% Spectral extension
    % left
    \node[minimum height=50, minimum width=10, anchor=south] (l-up) at (1.75*\dhor,-0.5*\dvert) {};
    \node[block, fill=\myBlue, minimum height=30, minimum width=10, anchor=south] () at (1.75*\dhor,-0.5*\dvert) {};

    \node[minimum height=50, minimum width=10, anchor=north] (l-do) at (1.75*\dhor,-0.5*\dvert) {};
    \node[block, fill=\myGreen, minimum height=30, minimum width=10, anchor=north] () at (1.75*\dhor,-0.5*\dvert) {};
    
    \node[block, fill=white!80!black, minimum height=40, minimum width=10] () at (1.75*\dhor,-0.5*\dvert) {};

    % right
    \node[minimum height=70, minimum width=10, anchor=south] (r-upup) at (2.25*\dhor,-0.5*\dvert) {};
    \node[block, fill=\myGreen, minimum height=40, minimum width=10, anchor=south] () at (2.25*\dhor,-0.5*\dvert) {};
    
    \node[minimum height=50, minimum width=10, anchor=south] (r-up) at (2.25*\dhor,-0.5*\dvert) {};
    \node[block, fill=\myBlue, minimum height=30, minimum width=10, anchor=south] () at (2.25*\dhor,-0.5*\dvert) {};

    \node[minimum height=70, minimum width=10, anchor=north] (r-dodo) at (2.25*\dhor,-0.5*\dvert) {};
    \node[block, fill=\myBlue, minimum height=40, minimum width=10, anchor=north] () at (2.25*\dhor,-0.5*\dvert) {};

    \node[minimum height=50, minimum width=10, anchor=north] (r-do) at (2.25*\dhor,-0.5*\dvert) {};
    \node[block, fill=\myGreen, minimum height=30, minimum width=10, anchor=north] () at (2.25*\dhor,-0.5*\dvert) {};
    
    \node[block, fill=white!80!black, minimum height=40, minimum width=10] () at (2.25*\dhor,-0.5*\dvert) {};

    % Arrows
    \path[edge, dashed, draw=\myBlue,triangle 45-triangle 45] (l-up) -- (r-up);
    \path[edge, dashed, draw=\myBlue,triangle 45-triangle 45] (l-up) --++ (0.9,0) |- (r-dodo);

    \path[edge, dotted, draw=\myGreen, triangle 45-triangle 45] (l-do) -- (r-do);
    \path[edge, dotted, draw=\myGreen, triangle 45-triangle 45] (l-do) --++ (1.1,0) |- (r-upup);

    \node[block, draw=white, anchor=east] () at (1.1*\dhor, -0.5*\dvert) {Spectral Extension:};

    \draw [ultra thick, decorate, decoration = {brace,mirror,amplitude=5pt}] (1.65*\dhor,-0.5*\dvert+1.1) --  (1.65*\dhor,-0.5*\dvert-1.1) node[block, draw=white, pos=0.5,left=0.2]{$\Ndata$};

    \draw [ultra thick, decorate, decoration = {brace,mirror,amplitude=5pt}] (1.65*\dhor,-0.5*\dvert+1.1) --  (1.65*\dhor,-0.5*\dvert-1.1) node[block, draw=white, pos=0.5,left=0.2]{$\Ndata$};
    \draw [ultra thick, decorate, decoration = {brace,amplitude=5pt}] (2.48*\dhor,-0.5*\dvert+1.4) --  (2.48*\dhor,-0.5*\dvert-1.4) node[block, draw=white, pos=0.5,right=0.2]{$\Nsc$};

    \node[text=\myGreen, anchor=west, font=\large] () at (2.28*\dhor,-0.5*\dvert+1.3) {$\Nse$};
    \node[text=\myBlue, anchor=west, font=\large] () at (2.28*\dhor,-0.5*\dvert-1.3) {$\Nse$};

\end{tikzpicture}
            \caption{
                Block diagram DFT-s-OFDM with SE and FDSS. 
                The SE operation is sketched in the middle.
            }
            \label{fig:system-model}
        \end{minipage}\hfill
        \begin{minipage}[b]{.38\textwidth}
            \begin{tikzpicture}
\begin{axis}[
    axis line style = {draw=none},
    tick style={draw=none},
    xtick = \empty,
    ytick = \empty,
    width=\textwidth,
    height=0.5\textwidth,
    ]

    \addplot+[line width=1pt, \myRed, solid, mark=none, domain=-10:-8]   {(x+8)^3 + (x+8)^2 + 1};
    \addplot+[line width=1pt, \myRed, dashed, mark=none, domain=-8:-6]   {-(x+8)^3 + (x+8)^2 + 1};
    \addplot+[line width=0.5pt, black, dashed, mark=none] table[x=X, y=Y, col sep=comma] {
X, Y
-8, -3
-8, 1.5
    };

    \addplot+[line width=1pt, \myBlue, solid, mark=none] table[x=X, y=Y, col sep=comma] {
X, Y
-2, 1
2, 1
    };
    \addplot+[line width=0.5pt, black, dashed, mark=none] table[x=X, y=Y, col sep=comma] {
X, Y
0, -3
0, 1.5
    };
    \addplot+[line width=1pt, \myRed, dashed, mark=none, domain=2:3]   {-4*(x-2)^2 + 1};
    \addplot+[line width=1pt, \myRed, solid, mark=none, domain=-3:-2] {-4*(x+2)^2 + 1};

    \addplot+[line width=1pt, \myRed, solid, mark=none, domain=6:7.05]   {cos(3*deg(x-7))};
    \addplot+[line width=1pt, \myGreen, solid, mark=none, domain=5:6]   {cos(3*deg(x-6))-2};

    \addplot+[line width=1pt, \myRed, dashed, mark=none, domain=8.95:10]   {cos(3*deg(x-9))};
    \addplot+[line width=1pt, \myGreen, dashed, mark=none, domain=10:11]   {cos(3*deg(x-10))-2};
    
    \addplot+[line width=1pt, \myBlue, solid, mark=none] table[x=X, y=Y, col sep=comma] {
X, Y
7, 1
9, 1
    };
    \addplot+[line width=0.5pt, black, dashed, mark=none] table[x=X, y=Y, col sep=comma] {
X, Y
8, -3
8, 1.5
    };

    \node[circle, fill=\myBlue, text=\myBlue, inner sep=1] () at (axis cs: 10,-1) {};
    \node[circle, fill=\myBlue, text=\myBlue, inner sep=1] () at (axis cs: 6,-1) {};

    \node[text=\myRed, anchor=east, fill=none, inner sep=1pt, font=\tiny] () at (axis cs: -10,1) {$F_k$};
    \node[text=\myRed, anchor=west, fill=none, inner sep=1pt, font=\tiny] () at (axis cs: -7.5,1.3) {$F_{\Nsc-k}$};

    \node[text=\myRed, anchor=west, fill=none, inner sep=1pt, font=\tiny] () at (axis cs: -4.7,0) {$F_k$};
   \node[text=\myBlue, anchor=south, fill=white, inner sep=0pt, font=\tiny] () at (axis cs: 0,1.2) {$C$};

    \node[text=\myRed, anchor=east, fill=none, inner sep=1pt, font=\tiny] () at (axis cs: 6.25,0.5) {$F_k$};
    \node[text=\myGreen, anchor=west, fill=none, inner sep=1pt, font=\tiny] () at (axis cs: 11,-2) {$\sqrt{C^2-F_{k}}$};
    \node[text=\myBlue, anchor=south, fill=white, inner sep=1pt, font=\tiny] () at (axis cs: 8.5,1.1) {$C_v=1/\sqrt{\Ndata}$};
    \node[text=\myBlue, anchor=west, fill=none, inner sep=1pt, font=\tiny] () at (axis cs: 10.5,-0.8) {$C_v/\sqrt{2}$};

    \addplot+[line width=0.5pt, white, dashed, mark=none] table[x=X, y=Y, col sep=comma] {
X, Y
11, -3.0
15, -3.0
    };

\node[anchor=south, fill=white, inner sep=1pt, font=\tiny] () at (axis cs: -8,-3.2) {(a)};
\node[anchor=south, fill=white, inner sep=1pt, font=\tiny] () at (axis cs: 0,-3.2) {(b)};
\node[anchor=south, fill=white, inner sep=1pt, font=\tiny] () at (axis cs: 8,-3.2) {(c)};

\end{axis}

\end{tikzpicture}
            \caption{
                Constrained FDSS designs: (a) non-flat; (b) flat; (c) flat with vestigial sideband (zero-ISI).
                The vertical dimension represents the filter values in frequency domain, while the horizontal axis (omitted) represents the subcarrier index.
            }
        \label{fig:FDSS-design}
        \end{minipage}
    \end{figure*}
    
    Consider the system model of Fig.~\ref{fig:system-model}, representing a DFT-s-OFDM system with SE and FDSS.
    Assume that the input of the system is a vector of $\Ndata$ indices $\vect{s} = [s_1,\dots,s_{\Ndata}] $, where $s_i\in\{1,\dots,M\}$.
    The indeces $\vect{s}$ are mapped to modulation symbols $\vect{x} = [x_1,\dots,x_{\Ndata}]$, where the symbols $x_i\in\mathcal{M}$ belong to a constellation $\mathcal{M}\subset\mathbb{C}$ of cardinality $M$.
    The modulation symbols go through DFT processing and the resulting frequency-domain symbols are denoted by $\vect{X} = [X_1,\dots, X_{\Ndata}]$. 
    Symmetric SE is then applied to the frequency-domain symbols $\vect{X}$, and the extended symbols are denoted as $\vect{X}^\text{ext} = [X_1^\text{ext},\dots, X_{\Nsc}^\text{ext}]$, where $\Nsc$ is the number of the subcarriers at the SE output.
    The spectral extension operation is represented in Fig.~\ref{fig:system-model}. 
    Note that, at the transmitter side, the first $\Nse$ symbols of $\vect{X}$ are copied to the tail of $\vect{X}^\text{ext}$; similarly, the last $\Nse$ symbols of $\vect{X}$ are copied to the head of $\vect{X}^\text{ext}$.
    Hence, $\vect{X}^\text{ext}=[X_{\Ndata-\Nse+1},\dots,X_{\Ndata},X_1,\dots,X_{\Ndata},X_1,\dots,X_{\Nse}]$.
    The total number of subcarriers after the spectral extension is $\Nsc = \Ndata + 2 \Nse$.
    
    The extended symbols $\vect{X}^\text{ext}$ are then filtered with the FDSS filter with frequency response $\vect{F} = [F_1,\dots, F_{\Nsc}]$. 
    Since this processing is performed in the frequency domain, this corresponds to an element-wise (Hadamard) product. 
    The filtered symbols after FDSS are $\vect{\tilde{X}} = \vect{X}^\text{ext} \odot \vect{F}$,
    where $\odot$ denotes the element-wise multiplication.
    The FDSS filter energy is defined as $E_\text{FDSS} = \sum_{k=1}^{\Nsc} |F_k|^2$.
    After FDSS, the filtered symbols $\vect{\tilde{X}}$ go through an IFFT block of size $\Nfft$.
    Appropriate zero-padding is added to the IFFT's input, since $\Nfft > \Nsc$.
    The transmitted waveform in the discrete-time domain can be defined as $\vect{\tilde{x}}=[\tilde{x}_1,\dots,\tilde{x}_{\Nfft}]$.
    
    Transmitter and receiver circuitry such as channel coding/decoding for data bits, cyclic prefix addition/removal, digital-to-analog and analog-to-digital converters, amplifiers, and antennas are not illustrated for brevity.

    %%%%%%%%%%%%%%%%%%%%%%%%%%%%%%%%%%%%%%%%%%%%%%%%%%%%%%%%%%
    \subsection{Conventional Receiver}
    
    The receiver (lower branch in Fig.~\ref{fig:system-model}) observes $\vect{\tilde{y}}$, where each element is  $\tilde{y}_n = \tilde{x}_n + z_n$, $z_n\sim\mathcal{CN}(0,\sigma^2)$ are complex-valued i.i.d. Gaussian random variables with variance $\sigma^2$, and $n=1,\dots,\Nfft$.
    The received samples $\vect{\tilde{y}}$ go through FFT processing, and the resulting frequency-domain representation is denoted by $\vect{\tilde{Y}}$, where corresponding scheduled sub-carriers are utilized and the rest of FFT output is discarded.
    The receiver FDSS-SE operations are split into three parts:
    \begin{enumerate}
        \item Matched FDSS filter: this operation is performed in frequency domain by computationally-efficient element-wise multiplication of conjugation of FDSS and FFT output, as $\vect{R} = \vect{\tilde{Y}} \odot \vect{F}^*$, where $\vect{F}^*$ denotes the complex conjugate of $\vect{F}$.
        \item SE combining: the SE parts of $\vect{R}$ are combined in $\vect{T}$, and the replicas are discarded. 
        In other words: $\vect{T}=[R_{\Nse+1}+R_{\Nsc-\Nse+1},\dots,R_{2 \Nse}+R_{\Nsc}, R_{2 \Nse + 1}, \dots, R_{\Nsc -2 \Nse}, R_{\Nsc -2 \Nse+1} + R_1, \dots, R_{\Nsc -\Nse} + R_{\Nse} ]$.
        \item FDSS-SE normalization: multiply by $1/|\vect{\hat{F}}|^2$, where $\vect{\hat{F}} = [F_{\Nse+1}+F_{\Nsc-\Nse+1},\dots,F_{2 \Nse}+F_{\Nsc}, F_{2 \Nse + 1}, \dots, F_{\Nsc -2 \Nse}, F_{\Nsc -2 \Nse +1} + F_1, \dots, F_{\Nsc -\Nse} + F_{\Nse} ]$
    \end{enumerate}
    The output after the FDSS-SE normalization is denoted by $\vect{\hat{Y}}=[\hat{Y}_1, \dots, \hat{Y}_{\Ndata}]$.
    Then, IDFT processing is applied to obtain the time-domain symbols $\vect{y}$, and the demodulation operation estimates the transmitted indices $\vect{\hat{s}}$.

    %%%%%%%%%%%%%%%%%%%%%%%%%%%%%%%%%%%%%%%%%%%%%%%%%%%%%%%%%%%%%%
    \subsection{Performance Metrics}

    The overhead introduced by the spectral extension determines the excess bandwidth of the system, defined as $\EBW = 2 \Nse/ \Ndata$; the bandwidth overhead is also measured in terms of extension ratio $\ER = 2 \Nse / \Nsc$.
    
    For each OFDM symbol, the PAPR is defined as $\PAPR = \{\max_{n=1,\dots,\Nfft} |\tilde{x}_n|^2\}/{P_\text{TX}}$,
    where $P_\text{TX} = \sum_{n=1,\dots,\Nfft} |\tilde{x}_n|^2 / \Nfft$ is the average transmitted power for the OFDM block.
    PAPR's complementary cumulative density function (CCDF) curves are often used when evaluating system performance in terms of PAPR. 
    Note that the CCDF is defined as $\text{CCDF}(a)=\mathrm{Pr}[\text{PAPR}>a]$.
    An example of a CCDF curve is in Fig~\ref{fig:PAPR}, where the curves on the left provide better performance (lower PAPR).
    
    The SER can be used as an end-to-end metric to evaluate the overall system performance.
    The main source of SER degradation is the presence of ISI.
    SER is defined as $\SER = \mathbb{E} d_\text{H}(\vect{s}, \vect{\hat{s}})$, where $d_\text{H}(\vect{s},\vect{\hat{s}}) = \sum_i \mathbbm{1}(s_i \neq \hat{s}_i)$ is the Hamming distance.

%%%%%%%%%%%%%%%%%%%%%
%%%%%%%%%%%%%%%%%%%%%
\section{FDSS Design with Machine Learning}
\label{sec:FDSS-design-ML}

    Consider the end-to-end system model of Fig.~\ref{fig:system-model}, which is explained in Section~\ref{sec:system-model}.
    We propose a machine learning-based approach to learn the FDSS filter taps $\vect{F}=[F_1,\dots,F_{\Nsc}]$.
    The indices $\vect{s}$ and the AWGN $\vect{z}$ are sampled from proper probability distributions, and the filter taps $\vect{F}$ are updated with stochastic gradient descent according to an appropriate loss function defined below.
    The goal is to determine $\vect{F}$ such that the following requirements are taken into account: 
    \begin{enumerate}
        \item SER requirement: the end-to-end SER is minimized;
        \item PAPR requirement: the transmitted waveform exhibits low PAPR;
        \item FDSS shape requirement: a flat response in the center-band of $\vect{F}$ is desirable;
        \item FDSS scalability: the FDSS filter should be expressed as a function of the subcarrier indices, that can be adjusted for different values of $(\Ndata,\Nse,\Nsc)$.
    \end{enumerate}
    We address this problem on three levels, as explained in the next subsections.

%%%%%%%%%%%%%%%%%%%%%
\subsection{Loss Function}
\label{sec:loss}
    
    In order to align the system design requirements and the optimization objective, we propose a versatile loss function that enables the learned filter $\vect{F}$ to manifest the desired characteristics.
    The loss function is defined as:
    \begin{equation}
        \mathcal{L} = \mathcal{E} + \lambda \mathcal{P} + \gamma \mathcal{S} 
        \label{eq:loss}
    \end{equation}
    where $\mathcal{E}$ represents the SER component, $\mathcal{P}$ represents the PAPR component, and $\mathcal{S}$ represents the FDSS shape component; $\lambda$ and $\gamma$ determine the tradeoff between the three loss components.
    The tradeoff coefficients are assumed to be non-negative, i.e., $\lambda,\gamma \geq 0$.
    For example, if $(\lambda\to\infty, \gamma\ll \lambda)$, then the learned FDSS $\vect{F}$ aims to reduce PAPR only, without optimizing the constraints on the SER and the FDSS shape.

\subsubsection{SER component $\mathcal{E}$}
      
    The SER requirement is modeled through the mean squared error (MSE) between the transmitted symbols $\vect{x}$ and the received signal $\vect{y}$, i.e., $\mathcal{E} = \mathbb{E}_{\vect{s},\vect{z}} ||\vect{x}-\vect{y}||^2$. 
    Cross-entropy loss between $\vect{s}$ and \emph{soft} estimates of $\vect{\hat{s}}$ was also considered: however, in our experiments, cross-entropy loss did not produce better results compared to MSE loss.
    Moreover, the disadvantage of using cross-entropy loss is that a neural network-based demodulator has to be added to the system, adding complexity during training time.

\subsubsection{PAPR component $\mathcal{P}$}    
    
    The PAPR contribution to the loss function is modeled through two features: (i) the PAPR per OFDM symbol in dB scale, and (ii) a novel metric called area under the CCDF (AUCCDF), which aims to capture desired characteristics of the PAPR distribution.
    Hence, the overall PAPR loss component is $\mathcal{P} = \mathbb{E}_{\vect{s},\vect{z}} [ \lambda_1 \text{PAPR}_\text{dB} + \lambda_2 \text{AUCCDF}]$, 
    where $\text{PAPR}_\text{dB} = 10 \log_{10} \text{PAPR}$, $\text{AUCCDF}=\sum_{i=1}^{N_\text{bins}} [1-\text{CDF}(\text{PAPR}_i)]$, and $(\lambda_1,\lambda_2)$ are coefficients to represent the importance of these two components in the tradeoff.
    $N_\text{bins}$ denotes the number of equidistant bins used for the CDF approximation for values included in [0, $\text{PAPR}_\text{max}$] dB.
    The AUCCDF metric is introduced to drive the optimization towards filters that exhibit PAPR CCDF curves leaning ``to the left''.
    In order to make AUCCDF a differentiable metric, we approximate the CDF with a sum of shifted \emph{sharp} sigmoid functions, where the shifts correspond to the edge of the bins for the empirical CDF.

\subsubsection{FDSS shape component $\mathcal{S}$}

    The spectral flatness measure (SFM) is used as a proxy for the FDSS shape loss component. 
    The SFM is defined as the ratio between geometric mean and arithmetic mean. 
    In this work, we define the SFM for the mid-band, i.e.,  $\text{SFM} = \{\sqrt[N]{\prod_{k=N_a}^{N_b} F_k}\}/\{\frac{1}{N} \sum_{k=N_a}^{N_b} F_k\}$, 
    where $N_a = 2\Nse+1$, $N_b = \Nsc - 2\Nse- 1$, and $N = \Nsc - 2\Nse$.
    Note that $0\leq \text{SFM} \leq 1$, and $\text{SFM}\to 1$ for flat spectrum.
    In our loss function, the SFM in dB scale is used, i.e., $\text{SFM}_\text{dB} = 10\log_{10} \text{SFM}\in[-\infty,0]$.
    Hence, the FDSS shape component is $\mathcal{S} = \text{SFM}_\text{dB}$.

%%%%%%%%%%%%%%%%%%%%%
\subsection{Learned Filter Taps Model}

    The FDSS filter should be represented as a function within a compact interval, facilitating easy resampling for the desired number of subcarriers.
    As an example, conventional baseline filters are typically defined in terms of trigonometric functions (e.g., cosine)~\cite{Nyquist-1928} and hyperbolic functions (e.g., hyperbolic secant)~\cite{BetterThanNyquist, ImprovedNyquist, NovelTwoParams}. 
    In our approach, we propose the use of polynomial approximations to model the FDSS filter coefficients.
    By selecting an appropriately high polynomial order, this approach offers the requisite expressiveness to capture the filter's behavior effectively.
    The model for the learned filter taps $\vect{F}=\{F_k\}_{k=1}^{\Nsc}$ is
    \begin{equation}
        F_k = \sum_{d=0}^{D} a_d [s(k)]^d , 
        \label{eq:Fk-poly}
    \end{equation}
    where $D$ is the order of the polynomial approximation, $a_d$ are the (learned) polynomial coefficients, and $s(k)$ are the support value representing each subcarrier. 
    For example, we can impose that the support values $s(k)$ are equally spaced over a pre-determined interval $[-1,1]$: if more (less) subcarrier are required, more (less) values of $F_k$ can be computed according to~\eqref{eq:Fk-poly}.
    Note that during training the coefficients $a_d$ will be updated with gradient descent according to the loss function in~\eqref{eq:loss}.

%%%%%%%%%%%%%%%%%%%%%
\subsection{Constrained FDSS Design}
\label{sec:constrained-FDSS-design}

    Real-valued pulses in time domain exhibit even symmetry in frequency domain.
    In order to guarantee even symmetry by design with the polynomial approximation from~\eqref{eq:Fk-poly}, we impose the following constraints: 
    (i) the support values $s(k)$ are symmetric, i.e., $s(k) = -s(\Nsc-k+1)$ for $k=1,\dots,\Nsc$; 
    (ii) the polynomial powers are even, i.e., $d=0,2,4,\dots,D$ and $D$ is even.  
    This construction guarantees even symmetry for the filter taps, i.e., $F_k = F_{\Nsc-k+1}$ for $k=1,\dots,\Nsc$.

    Moreover, we can impose further constraints (in addition to even symmetry and energy normalization) on the filter taps to satisfy some desired properties, like flatness in the passband, or vestigial symmetry in the sideband (zero-ISI condition).
    In particular, we refer to the following constrained FDSS designs --- also sketched in Fig.~\ref{fig:FDSS-design}. 
\subsubsection{Non-flat} 
    No additional constraints are set on the filter values $F_k$. 
    The number of degrees of freedom (independent filter values) for this design is $\Nsc/2$.
\subsubsection{Flat}
    The passband is constrained to be a constant value $C$, i.e., $F_{2\Nse+1} = \dots = F_{\Nsc-2\Nse} = C$.
    The number of degrees of freedom for this design is $2\Nse+1$.
    Note that to impose this constraint, we can set the support values such that $|s(2\Nse+1)|=\dots=|s(\Nsc - 2\Nse)|$: hence, according to~\eqref{eq:Fk-poly}, the filter taps $F_k$ will have the same values (flat filter) for the passband, since we are considering even powers only.
\subsubsection{Flat with vestigial sideband}
    Zero-ISI Nyquist condition is imposed by design (zero-ISI pulse), as sketched in Fig.~\ref{fig:FDSS-design}-(c).
    For this constrained design, the polynomial approximation is used only for the upper sideband (red part in Fig.~\ref{fig:FDSS-design}-(c)), so a shorter support vector $s(k)$ can be used. 
    These filter values can be modeled with both even and odd powers in the polynomial approximation~\eqref{eq:Fk-poly}. 
    Even symmetry for the filter is guaranteed by imposing $F_k = F_{\Nsc-k+1}$ for $k=1,\dots,\Nsc$.
    The passband and lower sideband values (blue and green parts in Fig.~\ref{fig:FDSS-design}-(c)) are deterministic functions of the upper sideband values, i.e.:
    \begin{itemize}
        \item passband values: this is a deterministic constant value determined by the energy normalization for each filter, i.e., $F_{2\Nse+1} = \dots = F_{\Nsc-2\Nse} = C_v = E_\text{FDSS}/\sqrt{\Ndata}$;
        \item upper sideband: these are the learned values that can be modeled with the polynomial approximation~\eqref{eq:Fk-poly}.
        The filter taps are such that $C_v/\sqrt{2} \leq F_{\Nse+1}, \dots, F_{2\Nse} \leq C_v$;
        \item lower sideband: these are deterministic values satisfying the Nyquist criterion for zero-ISI filters in frequency domain, i.e., $F_k = \sqrt{C_v^2-F^2_{2\Nse-k+1}}, k=1,\dots,\Nse$.
    \end{itemize}   
    The number of degrees of freedom for this design is $\Nse$.

%%%%%%%%%%%%%%%%%%%%%
%%%%%%%%%%%%%%%%%%%%%
\section{Results}
\label{sec:results}
          
    For our simulations, we use a configuration compatible with 3GPP recommendations~\cite{3GPP}: we consider 32 physical resource blocks (PRBs) after SE, corresponding to $\Nsc=384$ subcarriers. 
    Extension ratio of $1/8$ is considered, leading to $\Nse=24$, $\Ndata = 336$, and excess banwdith $\EBW=14.2\%$.
    The FFT size is $\Nfft=1024$.
    We assume that the indices $s_i$ are equiprobable and their cardinality is $M=4$; QPSK modulation is adopted with unitary power constraint $\mathbb{E} ||x_i||^2=1$.
    The FDSS filters are normalized to unit energy, i.e., $E_\text{FDSS}=1$.
    At both the transmitter and receiver side, the DFT/IDFT and FFT/IFFT blocks are normalized by $1/\sqrt{\Ndata}$ and $1/\sqrt{\Nfft}$, respectively.
    The signal-to-noise ratio (SNR) is defined as $E_s N_0 = E_b N_0 \log_2(M) \frac{\Ndata}{\Nfft}$, where $E_s = 1$ and $N_0=\sigma^2$.
    We consider the well-known root-raised-cosine (RRC) filter~\cite{Nyquist-1928}, as a performance baseline for our simulation results.
    The product of the transmitter RRC filter and the matched filter at the receiver results in the raised-cosine spectrum, which guarantees zero-ISI.
    RRC FDSS already provides lower PAPR compared to the vanilla DFT-s-OFDM.
    In our performance evaluation, we consider the PAPR gain (lower PAPR) achieved by our learned FDSS w.r.t. RRC FDSS, and the corresponding SNR degradation (larger required SNR for the same SER).

    For the learned FDSS filter, we assume the polynomial degree for the approximation is $D=10$.
    The choice of $D=10$ was made after observing that increasing $D$ was not improving the overall performance in our experiments.
    Adam optimizer is used for the gradient updates of the coefficients $a_d$ in the FDSS filter model~\eqref{eq:Fk-poly}.
    The training is performed over at least $10^6$ OFDM blocks. 
    The performance is evaluated over a separate test set of size $10^5$.
    Both training and test data consist of samples of indices $\vect{s}$ and realizations of the complex AWGN $\vect{z}$.
    For the AUCCDF and the CDF approximation, the edges of the bins are defined as $[0, 0.05, 0.1, \dots,10]$ dB, and the \emph{sharp} sigmoid is $q(x) = 1 / [1-\exp(-100 x)]$.
    Multiple combinations of the Lagrangian multipliers $(\lambda_0,\lambda_1,\gamma)$ were tried in order to explore different tradeoffs between SER, PAPR, and FDSS shape components in~\eqref{eq:loss}.

\subsection{Limiting Cases}    

    \begin{figure}[t]
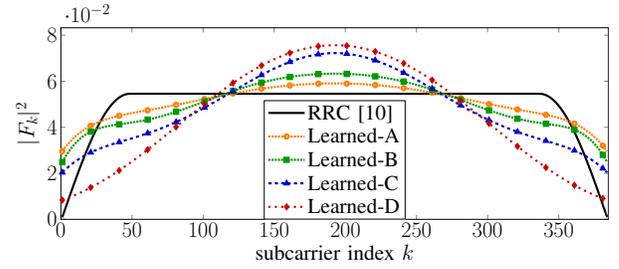

        \centering
        \includestandalone[width=0.9\columnwidth]{figures/fig_filters_limitingCases}
        \caption[FDSS filters for limiting cases]{
            Learned FDSS filters for limiting cases when $\EBW=14.2\%$.
            Resulting performance w.r.t. RRC FDSS: \vspace{-0.2cm}
        }
            \begin{center}
            \small
            \begin{tabular}{|c|c|c|}
            \hline
               Baseline: & SNR loss  & PAPR gain  \\ 
                RRC~\cite{Nyquist-1928} & at SER=$10^{-2}$ & at CCDF=$10^{-3}$ \\ \hline \hline
                Learned-A & $0.10$ dB & $1.15$ dB \\  \hline
                Learned-B & $0.25$ dB & $1.4$ dB \\  \hline 
                Learned-C & $1$ dB & $1.8$ dB  \\  \hline
                Learned-D & $3$ dB & $2.3$ dB  \\  \hline
            \end{tabular}
            \end{center}
        % }
        \label{fig:limit-filters}
    \end{figure}

    Fig.~\ref{fig:limit-filters} shows the learned FDSS filters when $(\lambda_1,\lambda_2) \gg \gamma$. 
    The Non-flat filter design is adopted for the results in this section.
    Increasing values of $(\lambda_1,\lambda_2)$ (i.e., increasing importance on PAPR in the optimization) are utilized in Learned-A/B/C/D, where case D gives the most importance to lowering the PAPR.
    The corresponding PAPR gains and SNR degradations (w.r.t. RRC FDSS filter) are reported in the caption. 
    Note that when $(\lambda_1,\lambda_2)$ are very large, then the optimization~\eqref{eq:loss} focuses on lowering the PAPR only; the resulting filters look like ``bell-shaped'' window functions. 
    In this case, one interpretation could be that it is beneficial from the PAPR point of view to spread subcarriers on different power levels.
    The drawback is substantial ISI which leads to large SNR degradation.

\subsection{``Almost Flat'' and Flat Filters}

    \begin{figure}[t]
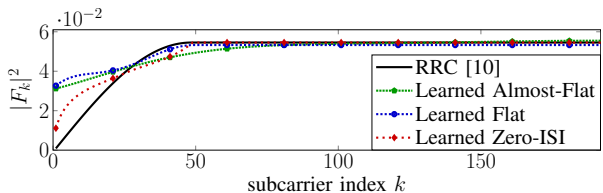

        \centering
        \includestandalone[width=0.9\columnwidth]{figures/fig_filters_main}
        \caption[Main result]{
            Learned FDSS filters with different constraints when $\EBW=14.2\%$.
            Only the first half of the subcarriers is shown in this plot.
            Resulting performance w.r.t. RRC FDSS:\vspace{-0.2cm}
        }
            \begin{center}
            \small
            \begin{tabular}{|c|c|c|}
            \hline
                Baseline: & SNR loss  & PAPR gain  \\ 
                RRC~\cite{Nyquist-1928} & at SER=$10^{-2}$ & at CCDF=$10^{-3}$ \\ \hline \hline
                Learned Almost Flat & $0.05$ dB & $0.8$ dB \\  \hline
                Learned Flat & $0.05$ dB & $0.65$ dB \\  \hline
                Learned Zero-ISI & $0$ dB & $0.5$ dB  \\  \hline
            \end{tabular}
            \end{center}
        % }
        \label{fig:FDSS}
    \end{figure}

   \begin{figure}[t]
        \centering
        \pgfplotsset{
    compat=newest,
    every mark/.append style={solid},
    /pgfplots/legend image code/.code={%
        \draw[mark repeat=2,mark phase=2,#1] 
            plot coordinates {
                (0cm,0cm) 
                (0.6cm,0cm)
                (1.2cm,0cm)%
            };
    },
}

\begin{tikzpicture}
\begin{axis}[
	scale only axis,
	font=\huge,
	width=\textwidth,
	height=0.4*\textwidth,
	xmajorgrids=true,
	ymajorgrids=true,
        yminorgrids=true,
	xmin=3,
	xmax=8,
        xtick={0,0.50,1,...,8},
	ymin=1e-4,
	ymax=1,
    ymode = log,
	xlabel={PAPR [dB]},
	ylabel={CCDF},
	legend columns=1,
	legend cell align={left},
        legend style={row sep=3pt, at={(0,0)}, anchor = south west},
	]

BINS,	noFDSS,	RRC,	AlmostFlat,	Flat,	ZeroISI
    \addplot+[line width=4pt, black, dashed, mark=none] table[x=BINS, y=noFDSS, col sep=comma] {figures/results_PAPR.txt};

    \addplot+[line width=4pt, black, solid, mark=none] table[x=BINS, y=RRC, col sep=comma] {figures/results_PAPR.txt};

    \addplot+[line width=4pt, \myGreen, densely dotted, mark=pentagon*, mark repeat=20, mark options={fill=\myGreen,solid}] table[x=BINS, y=AlmostFlat, col sep=comma] {figures/results_PAPR.txt};

    \addplot+[line width=4pt, \myBlue, dotted, mark=o, mark repeat=20, mark options={fill=\myBlue,solid}] table[x=BINS, y=Flat, col sep=comma] {figures/results_PAPR.txt};

    \addplot+[line width=4pt, \myRed, loosely dotted, mark=diamond*, mark repeat=20, mark options={fill=\myRed,solid}] table[x=BINS, y=ZeroISI, col sep=comma] {figures/results_PAPR.txt};

    \legend{
        Without FDSS-SE,
        RRC~\cite{Nyquist-1928},
        Learned Almost-Flat,
        Learned Flat,
        Learned Zero-ISI
    }
\end{axis}
\end{tikzpicture}
        \vspace{-0.2cm}
        \caption[PAPR]{
            CCDF of PAPR for the scenario in Fig.~\ref{fig:FDSS}.  
        }
        \label{fig:PAPR}
        \vspace{-0.5cm}
    \end{figure}
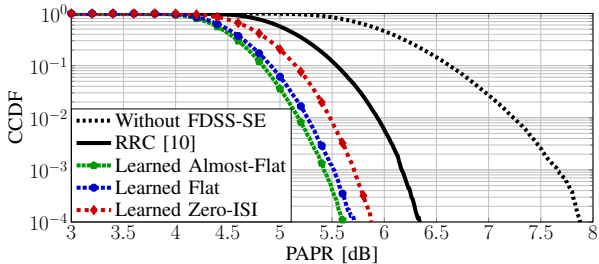

    Fig.~\ref{fig:FDSS} shows the FDSS filters learned with the different constrained designs explained in Section~\ref{sec:constrained-FDSS-design}.
    We select the FDSS filters that gave the lowest PAPR when the SNR degradation is $<0.05$ dB.
    For the ``Almost flat'' FDSS, the tradeoff between $(\lambda, \gamma)$ is chosen such that the learned filter resembles an impulse response as flat as possible in the passband.
    The resulting performance is listed in the caption.
    Note that imposing stricter constraints on the FDSS design results in reduced PAPR gains.
    The complete PAPR CCDF curves are shown in Fig.~\ref{fig:PAPR}.

\subsection{Resampled Filters vs Retrained}

    \begin{figure}[t]
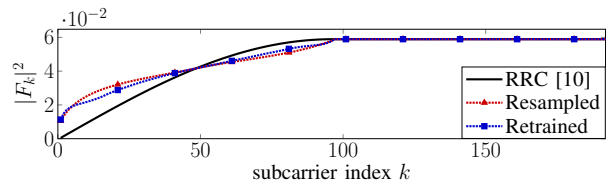

        \centering
        \includestandalone[width=0.9\columnwidth]{figures/fig_filters_resample}
        \caption[Resampling]{
            Learned FDSS filters with zero-ISI design when $\EBW=33.3\%$.
            The resampled filter is the same (learned zero-ISI for $\EBW=14.2\%$) as Fig.~\ref{fig:FDSS}, but with the sideband has been sampled more finely. 
            Performance w.r.t. RRC FDSS:
        }
            % \vspace{-0.2cm}
            \begin{center}
            \small
            \begin{tabular}{|c|c|c|}
            \hline
                Baseline: RRC~\cite{Nyquist-1928}   & PAPR gain at CCDF=$10^{-3}$ \\ 
                \hline \hline
                Resampled Zero-ISI (Fig.~\ref{fig:FDSS}) & $0.3$ dB \\  \hline
                Retrained Zero-ISI & $0.4$ dB \\  \hline
            \end{tabular}
            \end{center}
        % }
        \vspace{-0.5cm}
        \label{fig:resampling-retraining}
    \end{figure}

    In this section, we explore the robustness of the learned filters when \emph{reused} (resampled) for different $\EBW$ values.
    The original configuration is for $\Nsc=384$, $\Nse=24$ and $\EBW=14.2\%$; the resampled configuration is $\Nsc=384$, $\Nse=48$ and $\EBW=33.3\%$.
    Note that the number of subcarriers in the sideband is doubled for the latter case, i.e., twice the amount of points are sampled from the original polynomial approximation for the FDSS.
    For the sake of simplicity, only zero-ISI filters are considered for this experiment (which do not incur SNR degradation).
    The resulting FDSS filters are shown in Fig.~\ref{fig:resampling-retraining}.
    From this experiment, we can see that there is not a unique FDSS filter that provides the lowest PAPR values for all the values of $\EBW$.
    However, our end-to-end learned approach is able to outperform any baseline for all $\EBW$'s, and consequently provide a benchmark in terms of achievable PAPR gains in different system configurations.

%%%%%%%%%%%%%%%%%%%%%
%%%%%%%%%%%%%%%%%%%%%
\section{Conclusion}
\label{sec:conclusion}

    In this work, we presented a machine learning-based method to design FDSS filters in DFT-s-OFDM systems.
    Our proposed loss function captures the combination of SER, PAPR, and FDSS filter shapes.
    Different constrained designs have been considered to take into account the system necessities in terms of FDSS filter flatness or ISI requirements.
    In general, increasing FDSS design constraints can reduce or avoid SNR degradation for a reduced PAPR gain.
    The learned filters outperform conventional baselines in terms of the tradeoff between SER and PAPR, and our end-to-end approach can provide a useful benchmark to understand achievable PAPR gains in different system configurations.
    In particular, we have shown that the optimized filters differ for different EBW values. 
    We can learn filters for any EBW values, and then re-sample them for the appropriate number of subcarriers required by the system configuration.

\bibliographystyle{bibliography/IEEEtran} 
\bibliography{bibliography/references}

\end{document}